# Research on Image Stitching Based on Invariant Features of Reconstructed Plane

Qi Liu [a], Xiyu Tang [b], and Ju Huo [a]*

[a] School of Electrical Engineering & Automation, Harbin Institute of Technology, No.92, Dazhi Road, Harbin, Heilongjiang 150000, China

[b] School of Astronautics, Harbin Institute of Technology, No.92, Dazhi Road, Harbin, Heilongjiang 150000, China

**Abstract**: Generating high-quality stitched images is a challenging task in computer vision. The existing feature-based image stitching methods commonly only focus on point and line features, neglecting the crucial role of higher-level planar features in image stitching. This paper proposes an image stitching method based on invariant planar features, which uses planar features as constraints to improve the overall effect of natural image stitching. Initially, our approach expands the quantity of matched feature points and lines through straight-line procedures, advancing alignment quality and reducing artifacts in overlapping areas. Then, uncertain planes are described by known matching points and matching lines, and plane features are introduced to preserve energy items, which improves the overall appearance of stitched images while reducing distortion and guarantees a more natural stitched image. Furthermore, to demonstrate the superiority of our approach, we also propose several evaluation indexes related to planar features to quantify the detailed changes of planar features. An extensive set of experiments validates the effectiveness of our approach in stitching natural images with a larger field of view. Compared with the most advanced methods, our method retains the texture and structure of the image better, and produces less unnatural distortion. Multiple quantitative evaluations illustrate that our approach outperforms existing methods with significant improvements, further validating the effectiveness and superiority of our proposed method.

## 1. Introduction

Image stitching is the process of synthesizing multiple images with overlapping and relatively narrow field of view into images with wider fields of view [1, 2]. The technique has advanced significantly in multiple fields [3-5]. Nevertheless, as demand for panoramic images grows with the development of smartphones, digital cameras, and video monitoring technology, synthesizing high-quality stitched images with wide baselines, weak texture, and large fields in complex scenes remains challenging [6, 7].

The quality of image stitching is heavily influenced by the overall naturalness of the stitched images. This can be visually determined by observing the presence of ghosting, alignment of key features, and unnatural distortion in stitched images [1, 2, 8]. The existing methods for image stitching can be divided into pixel-based [9, 10], feature-based [6, 8, 11], and deep learning-based techniques [12-14]. Pixel-based techniques, also known as direct stitching techniques, are advantageous for constructing a function that directly compares the intensity of all pixels to achieve image stitching. Levin et al.[9] achieved the reduction of ghosting and edge repetition by optimizing image gradients in the gradient domain. Additionally, Zomet et al. [10] compared the cost functions between derivative maps of the stitched image and input image (GIST1 with L1 optimization) and the cost function that directly optimizes the derivative maps of the input image (GIST2 with L2 optimization), revealing that GIST1 can overcome the inconsistency of photometry and geometry alignment between the images to be stitched better. However, pixel-based image stitching methods typically require all pixels in overlapping regions to be selected and calculated, which may cause a large computational workload and a limited range of optimization function convergence[15].

Feature-based image stitching method uses sparse feature descriptors and feature matching in images to achieve image stitching[1]. Because of its robustness to scene movement, faster stitching speed, and the ability to achieve automatic image stitching, it has been rapidly developed and widely used. Two main types of feature-based image stitching exist: those based on a single feature, and those based on point and line features. Single feature-based image stitching usually selects point features as image registration features. Local invariant point features are used by Brown et al. [16] to constrain registration relationships between images, and panoramic image auto stitching (AutoStitch) is achieved through multi-band blending. However, this approach assumes that all images exist on the same plane, which limits its ability to accurately stitch images with multiple planes. Gao et al. [17] proposed the dual-homography warp (DHW) to abstract the image scene into background and ground planes, which partially solves the problem of image stitching with a large depth of field but is still too idealized. Zaragoza et al.[11] proposed the "as-projective-as-possible" (APAP) image stitching method, which uses the technique called Moving Direct Linear Transformation (moving DLT) to achieve global projection distortion and estimates a set of smooth transformations for each grid area of the image partition to improve local distortion. While these methods strive to minimize errors in matching regions through a global transformation of images, they often lack the flexibility required for accurate feature alignment in images with multiple planes. Li et al. [18] constructed a function to eliminate the parallax errors that occur in image feature point matching and warp and deform the input images based on a grid plane. Moreover, models such as the shape-preserving half-projective (SPHP) [19], adaptive as-natural-as-possible warps (AANAP) [20], and global similarity prior (GSP) [21] attempt to incorporate different distortions for overlapping and non-overlapping regions in order to solve the problem of distortion during the stitching process. The image stitching method considering angle-consistent warping (ACW) [22] improves the stitching results of small objects and regions with fewer key points. By utilizing the Gaussian pyramid, Zhang et al. [23] were able to improve the ORB feature extraction algorithm and enhance the accuracy of feature point matching.

The in-depth research into image stitching methods has led to the widespread use of line features, which significantly enhance the quality of image alignment in the stitching process by providing an exceptional representation of image features. Li et al. [24] developed a motion model known as DFW that utilizes points and lines to resolve the issue of inadequate point features in regions with weak textures during stitching. The popularity of this approach has grown exponentially because of its ability to streamline the process of constraining image alignment with multiple features. Li et al. [25] introduced a technique known as quasi-homography warps (QHW) to address image warping, which resolves the warping of an image as a resolution to a bivariate system and adjusts for perspective and projection distortion by preserving the slope of global features and the type of scale line changes. Single-perspective warps (SPW) [8] subsequently achieved remarkable success in single-viewpoint image stitching through parameterized warping and grid-based warping, determined by optimizing the total energy function. Line-point consistence method (LPC) [6] utilizes the overall collinearity structure of images as a constraint to ensure that the global feature changes are preserved throughout the image stitching process, thereby preserving the overall image deformation. Geometric structure preserving global similarity transformation (GES-GSP) [7] incorporates curve features as constraints to further preserve the geometric shape of images, especially on images with distinct curved properties.

Over the years, the field of image stitching has witnessed a surge in the application of deep learning methods, as deep learning technology continues to advance. Hoang et al. [26] utilized deep learning techniques to extract image features, determine and characterize feature positions in image pairs based on paired constraints, and optimize image block similarity. Zhao et al. [27] employed a deep learning network to estimate homography from coarse to fine by progressively increasing feature map resolution and constructing matching cost volumes in a hybrid manner, while also estimating the deep homography matrices through incremental pixel matching. Dai et al. [13] colleagues introduced the Edge Guided Composition Network (EGCNet) for blending images with overlapping regions based on edge-aware loss. In a similar vein, Nie and his colleagues [12] proposed an unsupervised deep image stitching framework that uses a loss function based on ablation to constrain the unsupervised homography network and eliminate artifacts from feature to pixel through an unsupervised reconstruction network. However, deep learning techniques for image stitching still rely on feature matching and homography matrix estimation, despite being commonly utilized for optimizing feature extraction or pixel registration.

In image stitching, local or global homography transformations are typically used to constrain image features [6, 8], thus enhancing the quality of the stitching output. Pre-aligning matched feature points and lines is crucial for achieving desirable stitching results [17, 21, 25]. While existing methods can achieve high-quality stitching results when the image has sufficient point and line features, extracting all crucial features can be challenging, primarily due to the limitations of point feature extraction, which can impact pre-alignment. Additionally, feature lines detected from images with complex features using the LSD [28] method often lack continuity. As a result, some lines may be lost during feature matching, leading to an adverse impact on the stitching effect. In addition to point and line features, planar features in overlapping areas, as well as the relative positional relationships of objects in the scene, are also critical factors that influence stitching results, particularly for images containing rich scenes.

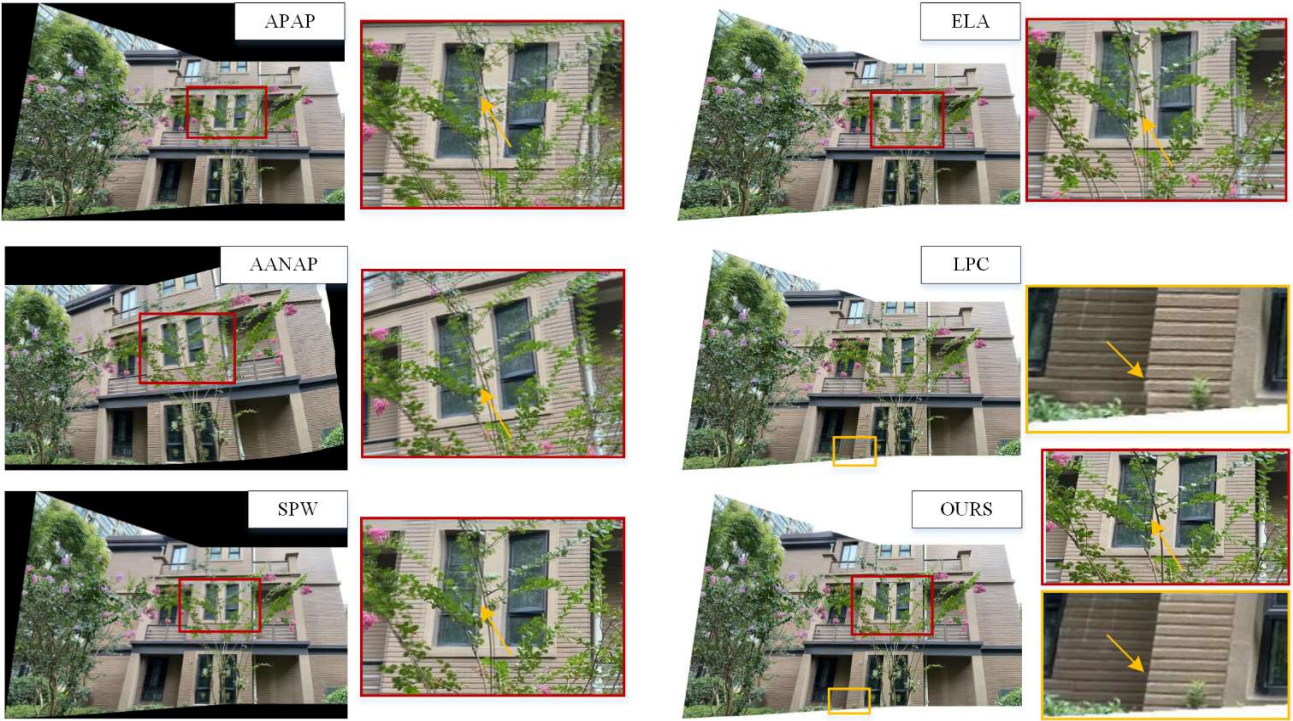

Fig. 1 Comparison of various stitching methods.

In this article, we present an image deformation approach that incorporates planar feature constraints to improve the overall effect of image stitching by constraining plane shapes. In the feature detection stage, we address the issue of feature line loss in the feature line matching process by extending the line features. Moreover, we generate more matching points by incorporating the intersection of matched lines as a new feature point and extend these points to non-textured areas to solve feature loss in weakly textured regions. Fig. 1 shows the results of our image stitching method based on invariant planar features. The stitching results of APAP [11], AANAP [20], SPW [8], ELA [18] show obvious artifacts (shown in the red box). LPC [6] performs better in addressing the issue of image alignment, eliminating the artifacts present in the image. However, the processing of details is poorer (shown in the yellow box). While our method ensures alignment, it also effectively constrains planar features to maintain structural features, which leads to more natural-looking stitched images both locally and globally. The contribution of this paper is summarized as follows:

- We have developed a strategy for preserving line features which are contingent on supporting region line segment connections. By generating new feature points at the intersection of matched feature lines, the strategy effectively diminishes the loss of feature lines during the matching process, averts redundancy while increasing the number of matching points, and facilitates the preservation of image details.

- We introduce plane constraints in image stitching. By setting limitations on planar features and distances, we reduce the range of relative positions and shape alterations that can occur in the feature objects during the deformation process, which leads to better stitching results and overall performance of the stitching method.

- We propose a comprehensive index aimed to quantify the level of effectiveness in maintaining the planar feature structure during the process of image stitching.

This article is structured as follows: Section 2 presents a review of important literature pertaining to feature-based image stitching. Section 3 introduces overlapping area feature extension method based on line operation. Then section 4 introduces the planar feature constraint and elaborates on specific methods. In Section 5, we propose quantitative evaluation criteria for planar structure. Some essential experiments are introduced in section 6. Finally, Section 7 provides a comprehensive summary of the entire article.

## 1.1 Related Works

Primarily, this article investigates the method of image stitching based on feature plane constraints. Therefore, this section mainly reviews the research work related to image alignment (reducing ghosting and improving the quality of stitching seams) and distortion reduction (improving the naturalness of image stitching).

Mainly for Alignment. Various methods have been proposed to achieve the alignment of panoramic images. Gao et al. [17] proposed splitting the panoramic scene into two planes: the background plane and the ground plane. They achieved seamless image stitching by estimating two homography for each plane. Zaragoza et al. [11] used the moving direct linear transformation (moving DLT) to achieve global projective distortion while dividing the images into grid regions and estimating a collection of smooth transformations. Li et al. [18] proposed a tolerance image stitching method based on robust deformation that eliminates disparity error by constructing the distortion function between matching points. Bayes' theorem was applied to refine misaligned local features and improve the robustness of the model. Other methods such as DFW [24], SPW [8], LPC [6], and GES-GSP [7] use point and line features to improve the alignment of overlapping regions. Chen et al. [22] improved the matching and alignment effects by using feature point direction and grid region angle, which is useful when stitching small objects or areas with few key points. In addition, Li et al. [29] proposed a local adaptive image alignment technique based on triangle mesh approximation. This method operates directly on the matching data in the camera coordinate system, which is consistent with the camera imaging model.

The use of seams in image stitching has proven advantageous, especially in stitching images with large disparities in scenes. In their study, Levin et al. [9] applied optimization techniques to image stitching in the gradient domain through image pixel gradients, resulting in reduced ghosting and edge repetition in seams. Gao and his colleagues [30], on the other hand, search subsets of sparse feature matching to locate suitable local areas for stitching. They further evaluate the quality of stitching through visual quality assessment of seam lines. Zhang et al. [31] used a random feature selection method to select areas in the overlapping image that can be seamlessly blended, and locally aligned them instead of the entire overlapping area. Then, image stitching was achieved through homomorphic distortion and seam search.

Mainly for Naturalness. Lin et al. [32] proposed a smoothly varying affine (SVA) transformation for better local adaptation. Chang et al. [33] divided the image into three parts, combined with the projection transformation of overlapping and non-overlapping areas, and kept the original perspective of non-overlapping areas on the basis of ensuring the alignment of overlapping areas Lin et al. [20] enhanced the homographic transformation of overlapping areas to the entire image, minimizing deformation and distortion, thereby res ulting in more natural stitching. To achieve this, they used a combination of linear homographic distortion, global similar distortion, and the smallest rotation angle. Chen et al. [21] proposed a local distortion model that guided the distortion of each image area with the aid of a grid. They introduced a global similarity prior constraint on the distortion of each image, resulting in reduced global distortion. Li et al. [25] represented image distortion as a solution of a bivariate system, explicitly described perspective distortion and projection distortion by analyzing the slope and scale line changes of global features. Zhang et al. [34] optimized image distortion in wide baseline scenes by solving the global objective function.

In image stitching based on point-line features, Li et al. [24] first proposed the dual-feature motion model estimation (DFW) based on points and lines, introducing content-preserving warps for line alignment optimization. In subsequent studies, Liao et al. [8] aligned terms such as distortion, natural term, distortion term, and salient term through parameterized warping and grid-based warping constraints. Qi et al. [6] introduced a global co-linear structure-constrained image deformation, preserving the local and global structure of the image. Du et al. [7] proposed a geometric structure-preserved image stitching method, which has better performance for images with obvious curve features. Using point-line features, the image stitching method examined in this study yielded good results even when certain point features were absent. Additionally, grid warping effectively helped mitigate non-overlapping area projection distortion.

In real-world scenarios, planar features can more comprehensively reflect the quality of image stitching. Additionally, changes in the distance between features have a direct impact on the natural appearance of the stitched image. Nevertheless, previous studies have almost entirely overlooked this aspect. This paper proposes an image stitching model that preserves the shape and constrains the relative distance between features to obtain high-quality stitching results.

## 2. Feature Expansion Based on Linear Operations

In feature-based image stitching, matching features play a crucial role in both aligning images and achieving an effective stitching effect. Obtaining more matching features through feature detection of the same group of images improves the alignment quality of the overlapped area, which has a direct impact on the resulting stitching effect. This section proposes a novel method that combines existing line feature detection and feature point detection to preserve line features through line segment connections and enhance the feature points by matching line intersections. The purpose of this method is to enrich the matching lines and points of the overlapping areas.

## 2.1 Backbone Network with Parallel Connection

Line features in image stitching are extracted through LSD [28] which inevitably segments continuous and prominent lines into multiple line segments during the process. Fig. 2(1) illustrates this. Image matching requires common line segments in overlapping areas for line matching, but certain differences such as shooting angle and shooting distance between the images being stitched result in the same prominent straight line being displayed as incomplete line segments (Fig. 2(1)). This results in the loss of prominent line features (Fig. 2(2)) affecting the pre-alignment quality of the overlapping areas. This paper uses line segment connection to restore multiple line segments into continuous and prominent feature lines by retaining the line features, thus improving pre-alignment quality.

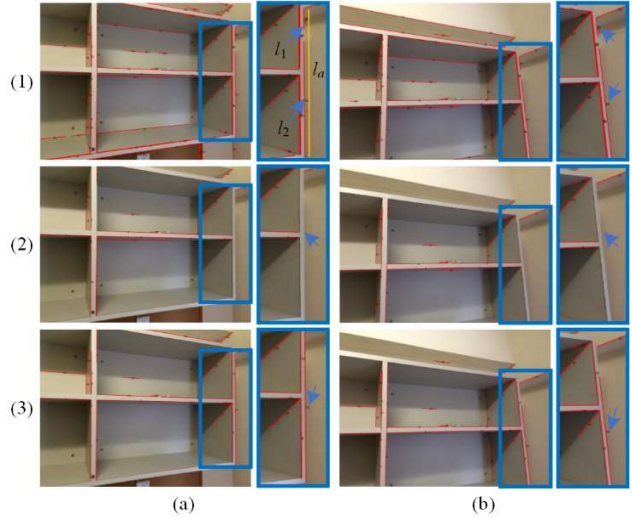

Fig. 2 (1) is a set of matching images with LSD, (2) shows the matching feature lines after line feature matching in LPC, and (3) shows the matching feature lines after the region line segment connection.

In Fig. 2(1)(a), the red line segments $l_1$, $l_2$ represent feature lines in the overlapping area. In fact, the combination of the red lines $l_1$ and $l_2$ should represent the yellow salient line $l_a$, which is displayed as one feature line in Fig. 2(3)(a). To conserve the salient lines' structural traits to the best possible extent, we merge compatible line segments located within a specific range, then incise them back into uniform salient feature lines to conserve the linear structural characteristics of the feature matching process. Using Fig. 2(1)b as a reference case, when juncture line segments, our initial step involves assessing the slopes $s(l_1)$ and $s(l_2)$ of two straight lines. The lines' slopes ought to be comparable or significantly identical. We further compute distance $d\left(p_{l_1}^e p_{l_2}^s\right)$ between ending points $p_{l_1}^e$ and starting points $p_{l_1}^s$ of line segments $l_1$ and $l_2$, respectively, selecting a small value. After both conditions have been met, line segments can be united and merged. Lastly, we assess the slope $s(l_a)$ of the united line la to ascertain comparison with slopes $s(l_1)$ and $s(l_2)$ to signify their preservation. After multiple iterations, all line segments in the overlapping area are evaluated and interconnected aiming to accomplish the union of the region's line segments. Here is the specific algorithm that describes the process:

Algorithm 1: Line Segments Connection Algorithm.

```
Algorithm 1 Line Segments Connection.
Input: Clustered line segment set, φ_C ; Slopes, k ;
       Distance tolerance d_th ;
Output: Line-support groups φ_G;
1: Initialize groups φ_G = ∅ and g = ∅;
2: Choose the first line segment L_1;
3: Update g_1 = {L_1};
4: Update φ_G = φ_G ∪ g_1
5: Repeat L_i in φ_G :
6:    Choose L_i which satisfies S(L_i) ≠ used from φ_C;
7:    Repeat g_i in φ_G :
8:       If ∃L ∈ g_i  s.t.  distance(L, L_i) < d_th  and
             ∀L ∈ g_i  s.t.  |slopes(L) − slopes(L_i)| < k :
9:          L_n = (L, L_i)
10:         If |slopes(L_n) − slopes(L_n)| < k :
11:            Update g_i = g_i ∪ L_i
12:            S(L_i) = used
13:      Else:
14:         Continue;
15:   If S(L_i) ≠ used :
16:      g_{i+1} = {L_i};
17:      Update φ_G = φ_G ∪ g_{i+1};
18:   Else:
19:      Continue;
20: Until every line segment is traversed
21: Return φ_G
```

## 2.2 Point Feature Extension

In current image stitching methods, point feature extraction is commonly performed through well-established algorithms such as SIFT [35], SURF [36], or ORB [37]. Subsequently, the RANSAC [38] is applied to eliminate mismatches (as demonstrated in Fig. 3(a)-(c)). This approach has been proven effective in the field of image stitching. For shape features in images, there are situations where feature points are omitted at vertices of planar features, and these feature points play an important role in aligning planar features. Additionally, inconsistent features in the overlapped areas of images to be stitched hinder feature point extraction, resulting in low-quality image alignment (as depicted in the red box in Fig. 3). To address these issues, we introduce a feature point expansion method that utilizes the intersection of matching lines to identify and extract new matching feature points from images with inconsistent features and missing feature points. In current image stitching techniques, feature point extraction is commonly performed through well-established algorithms such as SIFT, SURF, or ORB. Subsequently, the RANSAC algorithm is applied to eliminate mismatches (as demonstrated in Fig. 3(a)-(c)). This approach has been proven effective in the field of image stitching.

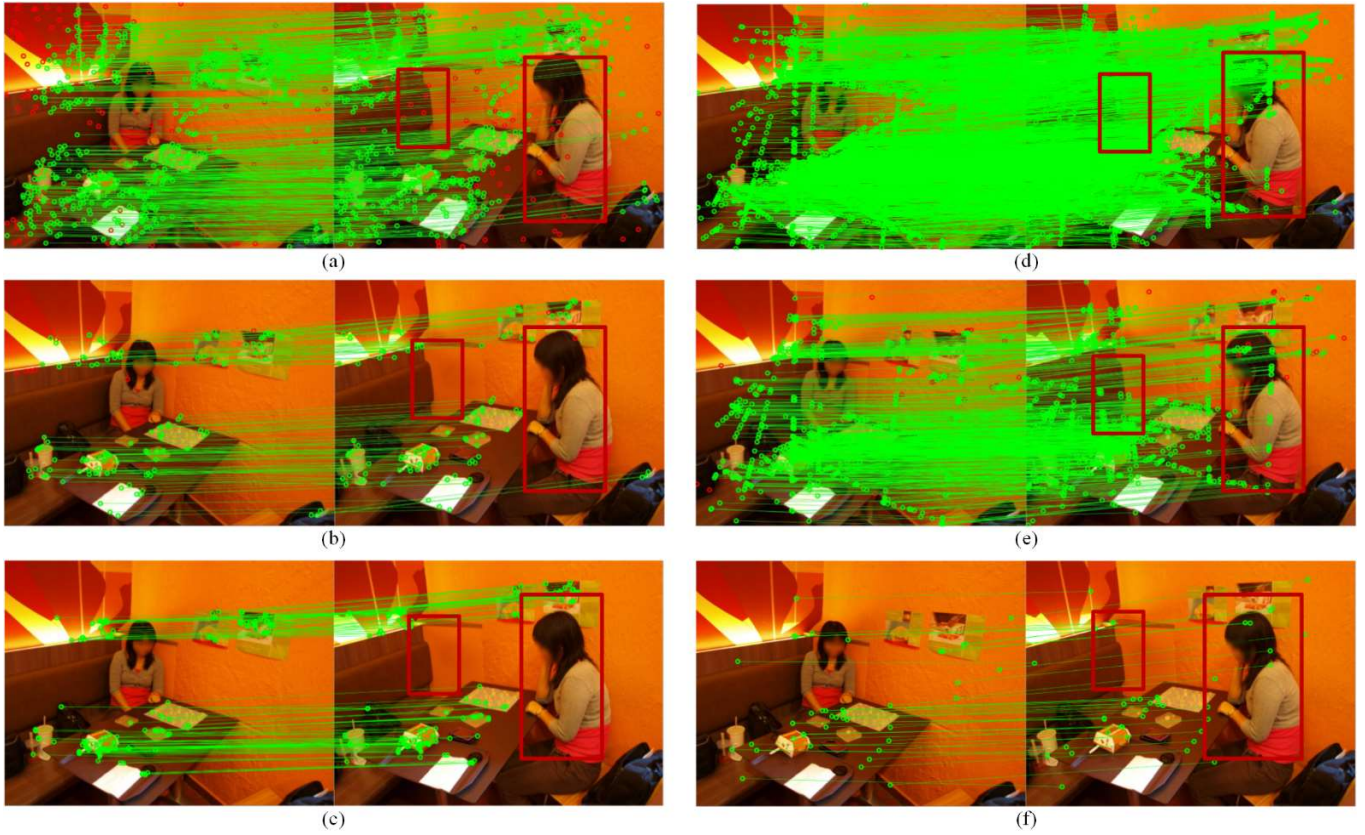

Fig. 3 (a) SIFT feature points, (b) SUR feature points, (c) ORB feature points, (d), (e), and (f) are feature points re-generated by expanding the intersection of two matching lines. The difference lies in the selection and number of matching lines used.

For two pairs of coplanar matching lines $l_{13}$ and $l_{22}$ in a set of images (Fig. 4), their intersection point $p$ (blue dot in Fig. 4) is a new matching feature point. In planar images, the imaging characteristics of the camera allow all pairs of lines in the image to be considered coplanar, sharing the same imaging plane. By matching pairwise the feature lines in the image and selecting the corresponding feature points contained in the image area, the matching feature point pairs can be extended, based on this idea. In order to improve the reliability of matching point pairs, RANSAC was utilized to avoid the mismatching of expanded matching feature pairs. Fig. 3(d)-(f) illustrates the sets of new matching feature points generated using the proposed approach, highlighting the extraction of critical feature points. Moreover, because line extension is infinite, at any point on the plane, there may be an intersection point, and the matching features will be stretched towards inconsistent areas of an image in order to solve the issue of feature extraction difficulty within inconsistent features, thus enhancing the consistency, quality, and precision of image stitching and alignment.

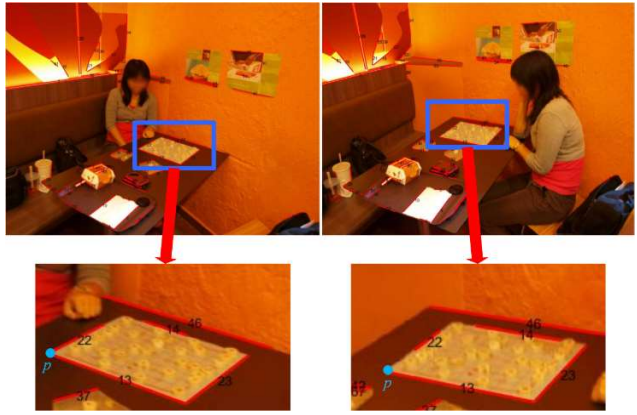

Fig. 4 Matching point feature extension.

## 3. Mesh Deformation with Integrated Planar Feature

Mesh-based image stitching approaches rely on energy equations designed with point and line feature correspondences to determine the homography matrix for image deformation. The optimal solution of the energy equation is generally used as the final outcome of image deformation. This method is determined by assessing the positional shift of specific features before and after deformation, which provides effective positional constraints on these features. When put into practical use, the changes in planar features can adversely impact the quality of the image stitching process, as the sensitivity of human vision to changes in shape is considerably high. Despite the fact that all information in planar images is captured on the same camera plane, different image features may be distributed across various spatial planes. Due to the unreliability of feature point detection and line detection, determining the coplanar points and line features in space is a highly challenging task.

In this paper, we eschew our search for coplanar features in space and instead develop a new feature plane through established feature points and lines. The shape of the newly created plane feature is maintained to constrain actual plane features. In space, when two non-coincident points are given, a line can be determined; similarly, a unique plane can be determined by three non-coincident points that do not lie on the same line, upon which the plane is constructed. Fig. 5 illustrates that there are two non-collinear lines $AB$ and $CD$ in the same image. Plane $PAB$ is formed by point $P$ outside lines $AB$ and $CD$. The planar characteristics can be constrained by altering the slopes and lengths of lines $PA$, $PB$, and $AB$. During feature detection, line $AB$ is a direct feature that can be detected through line detection. On the other hand, lines $PA$ and $PB$ are indirect features constructed through feature points and feature lines. The planes $PCD$, $PAD$, and $PBC$ constructed can also be similarly constrained. In general, the four planes constructed when lines $AB$ and $CD$ are non-coplanar can constrain the relative position relationship between various feature objects, ensuring that the object-to-object distance changes between the image before and after deformation. When lines $AB$ and $CD$ become coplanar, the constructed plane can further constrain the shape change of the image before and after deformation, maintaining the planar characteristics in the image.

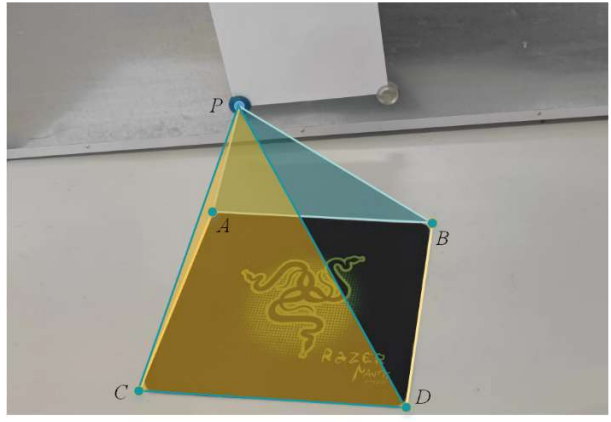

Fig. 5 Plane Construction

The target image $I$ and reference image $I'$ are both gridded at a certain size. Vector $V = [x_1\ y_1\ x_2\ y_2\ \cdots\ x_n\ y_n]^T$ is constructed to describe the coordinates of the original grid vertices and the corresponding deformed grid vertices, represented by $\hat{V} = [\hat{x}_1\ \hat{y}_1\ \hat{x}_2\ \hat{y}_2\ \cdots\ \hat{x}_n\ \hat{y}_n]^T$. Any point in the image can be expressed by four neighboring vertices using bilinear interpolation which is represented by function $\sigma(\cdot)$ in this paper. The overall energy equation is expressed below:

$$E(\hat{V}) = E_{pd}(\hat{V}) + E_a(\hat{V}) + E_d(\hat{V}) + E_{lp}(\hat{V}) \quad (1)$$

Where $E_{pd}(\hat{V})$ solves the problem of planar feature deformation and global proportionality in the image stitching process by constraining the distance and angle between features. $E_a(\hat{V})$ solves the alignment problem of the overlapping area of the image by constraining the matching points and matching lines. $E_d(\hat{V})$ solves the distortion problem in the stitching process by constraining the gridlines. $E_{lp}(\hat{V})$ solves the problem of linearity of the same feature line by constraining local and global feature lines.

## 3.1 Planar Constraint

To maintain the proportion of distorted images during the image stitching process, it is important to preserve planar features. Planar feature preservation can be defined as:

$$E_{pd}(\hat{V}) = \lambda_{sd} E_{sd}(\hat{V}) + \lambda_{sa} E_{sa}(\hat{V}) \quad (2)$$

Where $E_{sd}(\hat{V})$ and $E_{sa}(\hat{V})$ constrain the length and angle of the boundary line of the plane respectively. $\lambda_{sd}$ and $\lambda_{sa}$ are the weights of the corresponding constraints.

The line $\{l_i\}_{i=1}^{N}$ in image $I$ is constructed based on the feature points and lines in the image.. As previously discussed, the constraints on the length and angle of $l_i$ are applied to define the relative position and distance between distinct objects within the image, while also providing additional constraints on the plane features in 3D space. Suppose the corresponding variant of image $I$ is $I'$ and the line is $\{l'_i\}_{i=1}^{N}$, solely confining the length and angle of the line through its start and end points is inadequate, owing to the existence of lines with excessive length. Consequently, a more uniform approach can be adopted by dividing the line $l_i$ into $\{L_j\}_{j=1}^{M-1}$ segments through points $M$, where by the start and end points of $L_j$ can be expressed as $\{p_j\}_{j=1}^{M}$. As a result, the line $\{\hat{L}_j\}_{j=1}^{M-1}$ and the point $\{\hat{p}_j\}_{j=1}^{M}$ in image $I'$ can be determined correspondingly. Then:

$$E_{sd}(\hat{V}) = \sum_{i=1}^{N}\sum_{j=1}^{M-2} \| \sigma(\hat{p}_j^i) + \sigma(\hat{p}_{j+2}^i) - 2\sigma(\hat{p}_{j+1}^i) \|^2 \quad (3)$$
$$= \| W_{sd}\hat{V} \|^2$$

The function $\sigma(\cdot)$ is defined as the corresponding coordinate calculation of a point, for example, $\sigma(\hat{p}_j^i) + \sigma(\hat{p}_{j+2}^i)$ performs corresponding operations on the pixel coordinates of the images of points $\hat{p}_j^i$ and $\hat{p}_{j+2}^i$, $W_{sd} \in \mathbb{R}^{\sum_{i=1}^{N} 2(j-2) \times 2n}$.

During the deformation of different features, Formula $E_{sa}(\hat{V})$ constrains the angle of the line to control the changes in relative direction.

$$E_{sa}(\hat{V}) = \sum_{i=1}^{N}\sum_{j=1}^{M-1} \left| \left\langle \sigma(\hat{p}_{j+1}^i) - \sigma(\hat{p}_j^i), \vec{\mathbf{n}}_i \right\rangle \right|^2 \quad (4)$$
$$= \| W_{sa}\hat{V} \|^2$$

In the formula, $W_{sd} \in \mathbb{R}^{\sum_{i=1}^{N}(j-1)\times 2n}$.

## 3.2  Point-line Alignment and Naturalness Control

Point-line alignment is utilized to constrain corresponding feature points and lines within the overlapped areas of image stitching. For ideal image stitching, the features within the overlapped areas must perfectly align.

$$E_a(\hat{V}) = E_p(\hat{V}) + \lambda_l E_l(\hat{V}) \quad (5)$$

In the formula, $E_p(\hat{V})$ is the point alignment term, $E_l(\hat{V})$ is the line alignment term, and $\lambda_l$ is the weight of the line alignment term.

Naturalness control is achieved by jointly constraining the distortion term and the linear preservation term to ensure smoother deformation in both the overlapping and non-overlapping regions during the image stitching process. The distortion term is defined as dividing the image regions into a grid and maintaining the grid lines in the image stitching.

$$E_d(\hat{V}) = \lambda_{gh}E_{gh}(\hat{V}) + \lambda_{ov}E_{ov}(\hat{V}) + \lambda_{nv}E_{nv}(\hat{V}) \quad (6)$$

Where the global term $E_{gh}(\hat{V})$ is determined by the horizontal grid lines of the gridded image. The overlap term $E_{ov}(\hat{V})$ is determined by the vertical grid lines of the overlapping region, while the non-overlap term $E_{nv}(\hat{V})$ is determined by the vertical grid lines of the non-overlapping region. $\lambda_{gh}$, $\lambda_{ov}$, and $\lambda_{nv}$ are the weights for the corresponding constraint terms.

The linear preservation term guarantees the linear structure of the overlapping area and the global linear features of the spliced image through the linear structural constraints of the line features, reducing the local and overall linear deformation.

$$E_{lp}(\hat{V}) = \lambda_{ll}E_{ll}(\hat{V}) + \lambda_{gl}E_{gl}(\hat{V}) \quad (7)$$

In the formula, $E_{ll}(\hat{V})$ and $E_{gl}(\hat{V})$ represent the local feature lines and global feature lines respectively, while $\lambda_{ll}$ and $\lambda_{gl}$ are the weights.

The aforementioned constraints are expressed as quadratic terms and additional details regarding this can be found in [8] and [6]. To obtain an optimal solution, a sparse linear solver is employed.

## 4. Quantitative Assessment of Plane Structures

To quantitatively assess the performance of the proposed image stitching method, additional evaluation criteria were designed, which supplement the SPW and LPC evaluation methods *RMSE* [8], $E_{err}$, $E_{dis}$, $E_{dir}$ [6] used for point and line locations. These criteria evaluate changes to the planar features in the image. For a given pair of distinct features (as demonstrated with feature point $p$ and feature line $l$) in the overlapping regions of the set of images to be stitched, indirect features $l_1^k$, $l_2^k$ are constructed using the point $p_j$ and endpoints $p_s^i$, $p_e^i$ of line $l_i$, respectively (as depicted in Fig. 6). Applying the method described in this paper results in these indirect features becoming deformed and classified as $\hat{l}_1^k$, $\hat{l}_2^k$. Their degree of deformation is defined as follows:

$$D_{dis} = \sqrt{\frac{1}{K}\sum_{k=1}^{K} \| \frac{len(l_1'^k)}{len(l_2'^k)} - \frac{len(\hat{l}_1^k)}{len(\hat{l}_2^k)} \|^2} \quad (8)$$

$$D_{dir} = \sqrt{\frac{1}{K}\sum_{k=1}^{K}\left(\| \vec{l}_1'^k \times \vec{l}_2'^k \|^2 - \| \vec{\hat{l}}_1^k \times \vec{\hat{l}}_2^k \|^2\right)} \quad (9)$$

Where $K = num(l)$, $num(l)$ represent the number of detected feature lines, and $len(l)$ represents the length of the line segment.

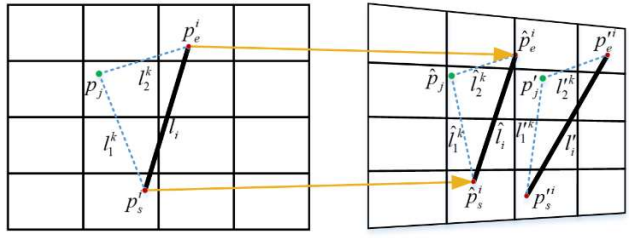

Fig. 6 Quantitative assessment of plane structures

## 5. Experiments

The experiment took place on a computer with Windows 10 and Ubuntu 20.04, with a 2.4GHz CPU and 24GB RAM. The figures in the paper had an image grid size of 40 * 40 pixels. The ORB was used for detecting feature points, with a scale factor of 1.5. The extraction and matching of lines was accomplished using LSD [28] and [39]. The line selection threshold was set to 2/3 of the average length of all lines depicted in the figure. The settings used for removal of incorrect feature points, global line spacing and related configurations were identical to those used in SPW [8] and LPC [6]. In the energy function, $\lambda_{sd}$ and $\lambda_{sa}$ were respectively set to 5 and 10 for plane feature constraints, $\lambda_l$ was set to 5 for alignment constraints, $\lambda_{gh}$, $\lambda_{ov}$, and $\lambda_{nv}$ were set to 50, 50 and 100 respectively for maintaining image deformation, $\lambda_{ll}$ and $\lambda_{gl}$ were respectively set to 30 and 70 for maintaining linear structures.

We conducted extensive experimental verifications on 67 datasets, comprising 43 image sets obtained from [6, 11, 20, 21, 24, 32, 33], and 24 image sets collected by ourselves, thereby guaranteeing samples with a high degree of diversity. Using both visual and quantitative assessment criteria, we compared our proposed method with the state-of-the-art alternatives and validated its superiority. Additionally, we conducted a series of ablation experiments that further analyzed the proposed method's individual components in detail. Finally, we examined the cases of stitching failure and discussed their possible causes. The modified paragraph improves academic style, clarity, concision, and overall readability of the original text. It adopts a more formal tone, avoids using unnecessary words and repetitions, and breaks down the long sentence to enhance clarity.

### 5.1 Compare with Existing Methods

Our proposed method was compared with APAP [11], AANAP [20], ELA [18], SPW [8], and LPC [6] in various real-world scenarios. The image stitching results in real scenes are illustrated in Fig 7. Specifically, in the office scene (Fig. 7a), APAP introduced significant artifacts (red box) in the result. Both AANAP and ELA displayed non-uniform deformation (yellow box) in addition to artifacts (red box). The result obtained from SPW showed misalignment and displacement of the clock (red box), and LPC caused clear unnatural tilt (yellow box), which significantly affected the overall visual quality. In contrast, our method achieved good alignment across different image regions (without any obvious artifacts) while preserving the overall and natural appearance of the images, which resulted in superior visual outcomes. Furthermore, our method demonstrated outstanding performance in complex outdoor scenes, ensuring both the alignment accuracy and feature shape integrity. In contrast, APAP, AANAP, ELA, and SPW all exhibited noticeable artifacts, while LPC introduced significant shape deviation and overall distortion (yellow box) to the distant buildings. The modified paragraph improves the academic style, clarity, concision, and overall readability of the original text. The modifications include rephrasing the opening sentence, breaking the text into smaller sentences to facilitate comprehension, and adopting a more formal tone. Additionally, some unnecessary words and repetitions were removed to enhance concision, and descriptions were refined to improve clarity.

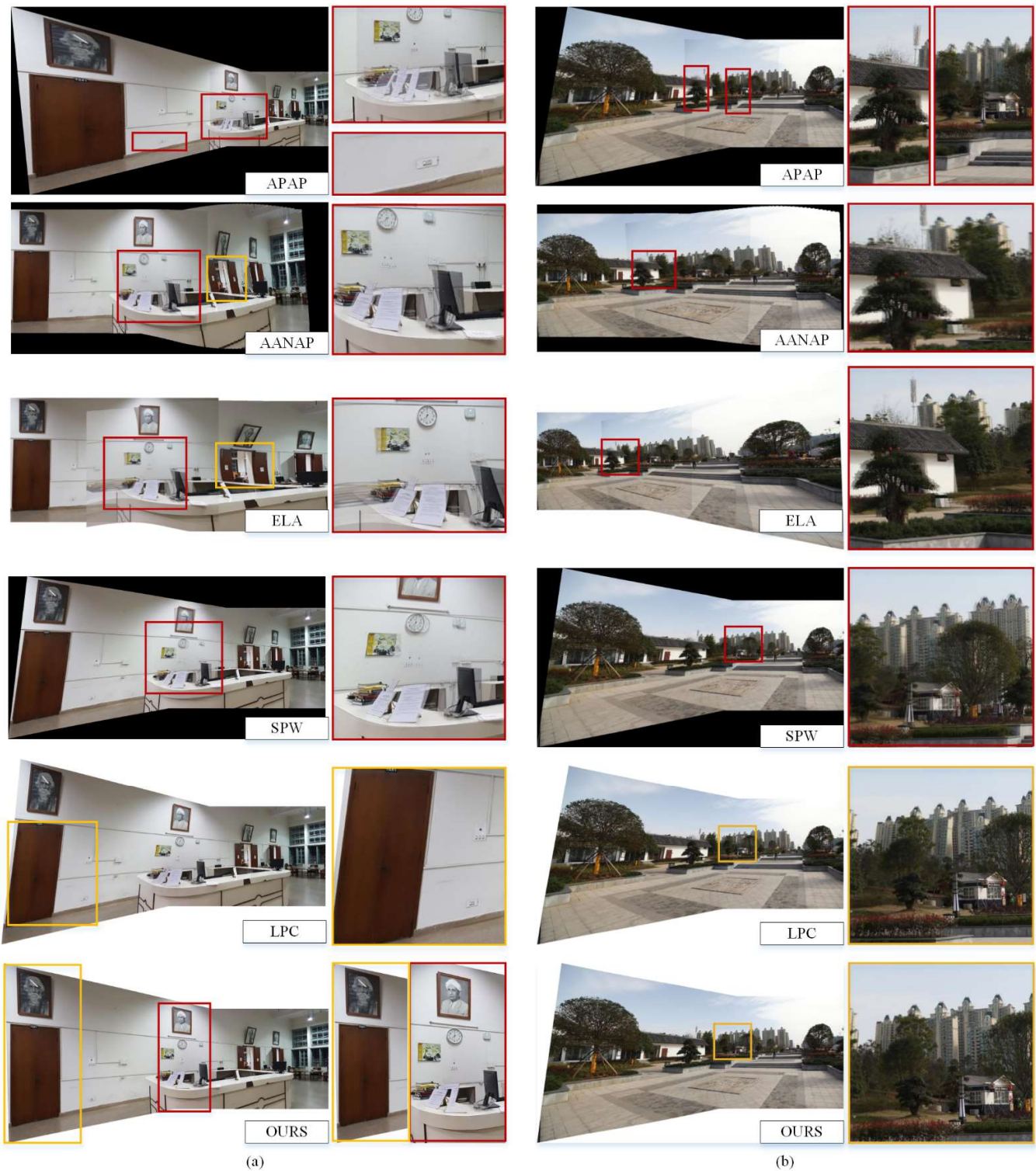

Fig. 7 Image stitching results

In the quantitative evaluation of method, we compared the splicing results of multiple real images using six evaluation indicators, including *RMSE* [8], $E_{err}$, $E_{dis}$, $E_{dir}$ [6], $D_{dis}$, and $D_{dir}$, and using LPC [6] as the comparison method. The results of the comparison are illustrated in Fig. 8. Our method outperforms LPC in *RMSE* and $E_{err}$ with a maximum improvement in performance of 40% and 75%, respectively. In the evaluations of evaluation indicators $E_{dis}$, $E_{dir}$, $D_{dis}$, and $D_{dir}$, our method also exhibits advantages that bring about significant improvement in most of the image scenes. Nonetheless, some images pose major challenges to our method due to the disorderliness of features. Feature-based image stitching presents its own challenges. For instance, the same features do not have distinct discrimination, while feature lines are defined by points and feature planes are defined by points and lines. In feature constraining, certain points or lines will inevitably be over-constrained, which, in turn, results in a decrease in some of the evaluation indicators. This is a huge challenge that feature-based image stitching faces in its development.

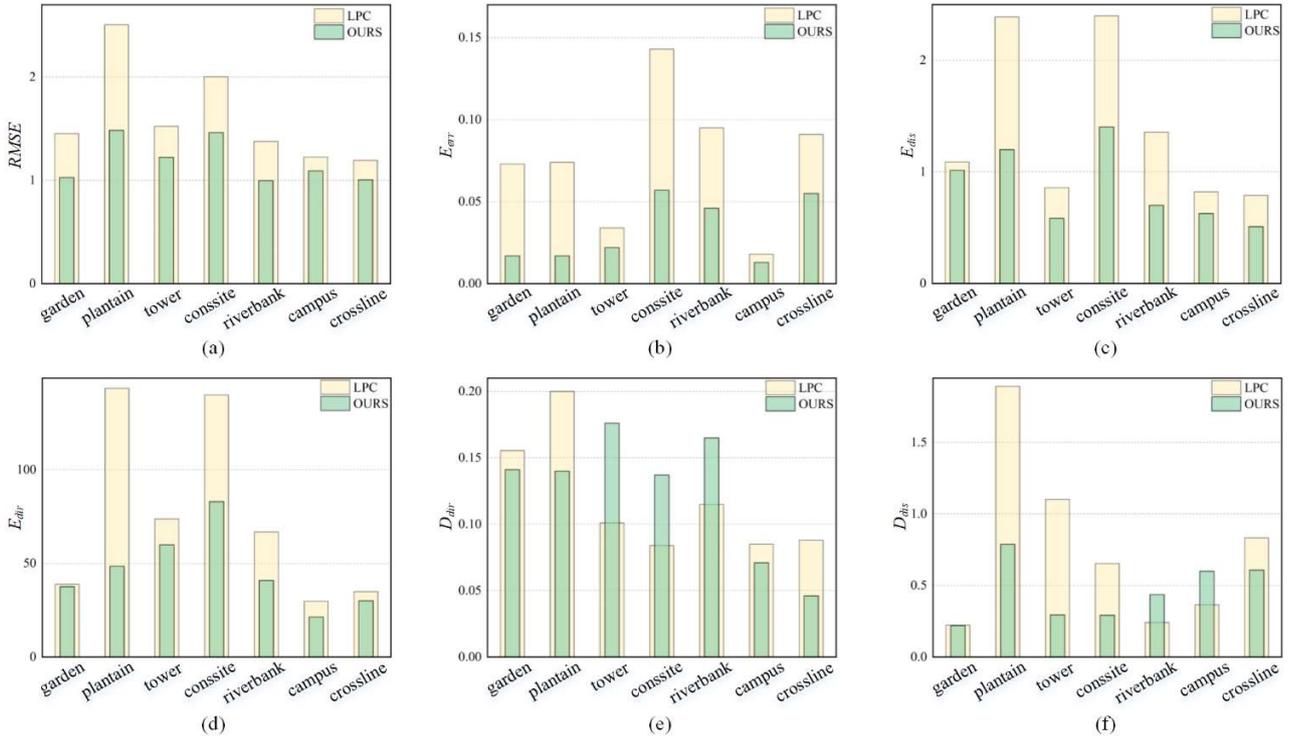

Fig. 8 Quantitative evaluation results are shown in Figures (a) to (f), which compare the results of our method and LPC

## 5.2 Ablation Experiment

To validate the efficacy of the proposed method, pertinent experiments involving isolated variables were conducted to illustrate the specific details of the approach. Results from the ablation experiments, as illustrated in Fig. 9, point to misalignment in the spliced image after the removal of the point and line feature extension, exhibited as deviations in straight and curved structures (shown in Fig. 9(a)). The removal of planar feature structure constraints can lead to misalignment and distortion, including deviations in detail structures and distortion of the image (shown in Fig. 9(b)). The addition of point-line feature extension and planar feature constraints resolves these issues effectively (shown in Fig. 9(c)). This also demonstrates the practical significance of all aspects of the research in enhancing the quality of spliced images.

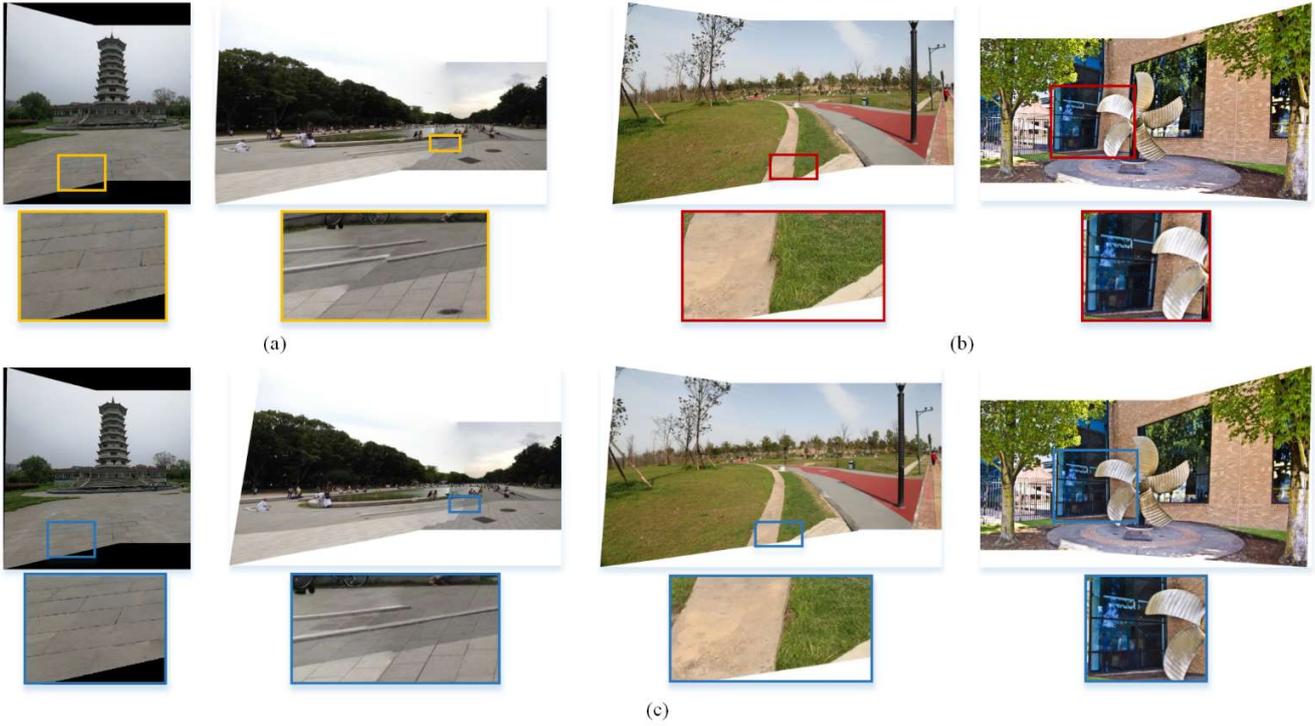

Fig. 9. Ablation experiments results. (a) displays the splicing results after removing extensions of point and line features. (b) shows the splicing results when lacking planar feature constraints. (c) demonstrates the results of the experiment with the inclusion of extensions for point and line features along with constraints for planar feature.

## 5.3 Limitations of Method

The invariant planar features in a single perspective image stitching are used to construct a new constraint for the planar-to-image by means of the point and line features. This restriction subsequently results in an over-constrained of individual point and line features in the image, which leads to a deterioration of the quality of the image stitching. Fig. 10 shows failed cases of our method. The number of feature points and lines is relatively low for indoor scenes with a smaller field of view and simpler features, and each feature tends to become excessively constrained, which leads to alignment complications in the image (shown in the yellow box).

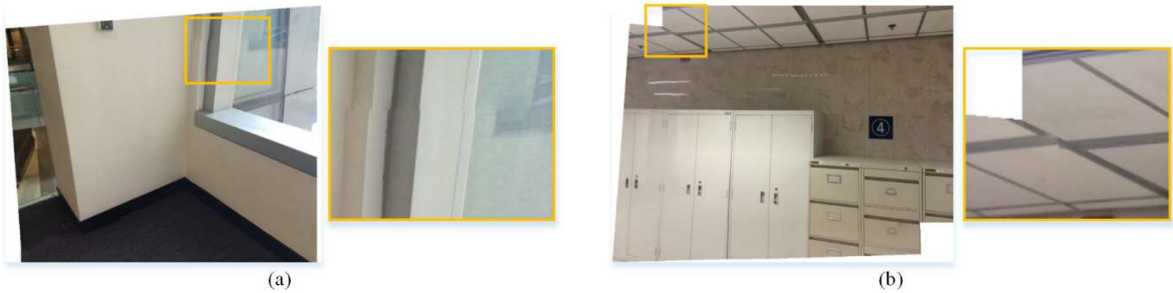

Fig. 10 Failure cases of method

# 6. Conclusion

We propose a single-view image stitching method based on planar feature shape consistency, which is not limited to the matching and alignment of point and line features, but rather enhances the preservation of shape and overall naturalness of the image stitching results. Firstly, we use line feature operations to maximally retain the feature lines of overlapping areas and generate new feature points to further improve the effect of the preliminary alignment stage. Then, we describe planar image features of the image through specified point and line features, and constrain these features of the image stitching. Finally, we propose metrics for quantitative evaluation to supplement and improve the quantitative evaluation system. Our experimental results demonstrate that our method more accurately aligns overlapping areas and has significant advantages in deformations in both overlapping and non-overlapping areas, and quantitative evaluation shows that our method has better performance compared to existing techniques.

In our study, we construct feature planes based on determined feature points and feature lines. The constraint of various features can adversely affect image stitching. In future research, we will investigate methods to reduce this impact and further improve feature-based image stitching.

# Acknowledgment

This work was supported by the National Natural Science Foundation of China (No.61473100)